\documentclass[showpacs,preprintnumbers,twocolumn,
amsmath,amssymb,groupedaddress,superscriptaddress]{revtex4}
\usepackage{graphicx}
\usepackage{dcolumn}
\usepackage{bm}

\begin{document}

\title{Calculations of $^{8}$He+p Elastic Cross Sections Using
Microscopic Optical Potential}

\author{V. K. Lukyanov}
\affiliation{Joint Institute for Nuclear Research, Dubna 141980,
Russia}

\author{E. V. Zemlyanaya}
\affiliation{Joint Institute for Nuclear Research, Dubna 141980,
Russia}

\author{K. V. Lukyanov}
\affiliation{Joint Institute for Nuclear Research, Dubna 141980,
Russia}

\author{D.~N.~Kadrev}
\affiliation{Institute for Nuclear Research and Nuclear Energy,
Bulgarian Academy of Sciences, Sofia 1784, Bulgaria}

\author{A. N. Antonov}
\affiliation{Institute for Nuclear Research and Nuclear Energy,
Bulgarian Academy of Sciences, Sofia 1784, Bulgaria}

\author{M. K. Gaidarov}
\affiliation{Institute for Nuclear Research and Nuclear Energy,
Bulgarian Academy of Sciences, Sofia 1784, Bulgaria}

\author{S. E. Massen}
\affiliation{Department of Theoretical Physics, Aristotle
University of Thessaloniki, 54124 Thessaloniki, Greece}

\begin{abstract}
An approach to calculate microscopic optical potential (OP) with
the real part obtained by a folding procedure and with the
imaginary part inherent in the high-energy approximation (HEA) is
applied to study the $^8$He+p elastic scattering data at energies
of tens of MeV/nucleon (MeV/N). The neutron and proton density
distributions obtained in different models for $^{8}$He are
utilized in the calculations of the differential cross sections.
The role of the spin-orbit potential is studied. Comparison of the
calculations with the available experimental data on the elastic
scattering differential cross sections at beam energies of 15.7,
26.25, 32, 66 and 73 MeV/N is performed. The problem of the
ambiguities of the depths of each component of the optical
potential is considered by means of the imposed physical criterion
related to the known behavior of the volume integrals as functions
of the incident energy. It is shown also that the role of the
surface absorption is rather important, in particular for the
lowest incident energies (e.g., 15.7 and 26.25 MeV/nucleon).
\end{abstract}

\pacs{24.10.Ht, 25.60.-t, 21.30.-x, 21.10.Gv}

\maketitle

\section[]{INTRODUCTION\label{s:intro}}

The experiments with intensive secondary radioactive nuclear beams
have made it possible to investigate the structure of light nuclei
near the neutron and proton drip lines as well as the mechanism of
scattering of the weakly bound nuclei. A special attention has
been paid to the neutron-rich isotopes of helium ($^{6,8}$He),
lithium ($^{11}$Li), berilium ($^{14}$Be) and others, in which
several neutrons are situated in the far extended nuclear
periphery and form a "halo". A widely used way to study the
structure of exotic nuclei is to analyze their elastic scattering
on protons or nuclear targets at different energies. Here we would
like to mention, for example, the experiments on scattering of
helium isotopes on protons at incident energies $E_{inc}$ less
than 100 MeV/N, namely, for $^{6}$He at energy 25.2
\cite{Ter99,Ter2000,Wolski98,Giot2004,Rusek2001}, 38.3
\cite{Lapoux2001}, 41.6 \cite{Cortina97,Lagoyannis2001,Cortina96}
and 71 MeV/N \cite{Korsh97a,Korsh97b}, for $^{8}$He at energy 15.7
\cite{Skaza2005}, 25.2 \cite{Ter2000}, 32
\cite{Korsh97a,Korsh97b}, 66 \cite{Korsh97a,Korsh97b} and 73 MeV/N
\cite{Korsh97a,Korsh97b,Korsh93} and also at energy 700 MeV/N for
He and Li isotopes (e.g.
\cite{Alkhazov2002,Egelhof2001,Egelhof2002,Neum2002,Egelhof2003}).

The experimental data on differential and total reaction cross
sections of processes with light exotic nuclei have been analyzed
using a variety of phenomenological and microscopic methods (e.g.
Refs.~\cite{Korsh97a,Korsh97b,Alkhazov2002,Egelhof2001,Egelhof2002,Neum2002,Egelhof2003,
Chulkov,Zhukov93,Zhukov94,Avrigeanu2000,Avrigeanu2002,Avrigeanu2001,Dortmans98,
Kara97,Deb2005,Amos2006,Deb2001,Deb2003,Amos2000a,Crespo95,Johnson97,Christley97,Crespo99,
Summers2002,AlKhal2007,Rodri2007,Arellano,Lukyanov2007}. Among the
latter methods we note, e.g. the microscopic analysis based on the
coordinate-space $g$-matrix folding method
\cite{Dortmans98,Kara97,Amos2006,Deb2005,Deb2001,Deb2003,Amos2000a,Arellano},
as well as works where the real part of OP is microscopically
calculated using the folding approach (e.g.
\cite{Avrigeanu2000,Avrigeanu2002,Avrigeanu2001,Dortmans98,Kara97,Lukyanov2007,
Satchler79,Khoa93,Khoa2000,Kara2000,Amos2000b}). Usually the
imaginary part of the OP's and the spin-orbit (SO) terms have been
determined phenomenologically. Thus, the OP's have a number of
fitting parameters. For example, OP's have been used to elaborate
the elastic differential cross sections of $^{6}$He+p,
$^{6}$He+$^{4}$He ($E_{inc}$=25 MeV/N) \cite{Avrigeanu2000} and
$^{6}$He+p and $^{8}$He+p ($E_{inc}<100$ MeV/N)
\cite{Avrigeanu2002} by means of the M3Y-Paris effective NN
interaction \cite{Khoa93,Khoa2000,Anant83}. In the calculations
the proton and neutron densities of the helium isotopes obtained
by Tanihata {\it et al.} \cite{Tani92} and also in the
Cluster-Orbital Shell-Model Approximation (COSMA)
\cite{Korsh97a,Korsh97b,Zhukov93,Zhukov94} were applied. It was
shown \cite{Avrigeanu2002} that the elastic scattering is
sensitive to different density distributions used in the folding
approach.

In  our previous work \cite{Lukyanov2007} in order to exclude the
usage of the phenomenological imaginary part of OP we have
performed calculations of $^{6}$He+p elastic differential cross
sections by means of the microscopic OP with the imaginary part
taken from the OP derived in \cite{Lukyanov2004,Shukla2003} on the
basis of the HEA \cite{Glauber,Sitenko,Czyz69}. This method
(Glauber approach) in its optical limit \cite{Czyz69} makes it
possible to obtain an analytic expression of the scattering
amplitude with the eikonal phase in the form of the so-called
profile function. The latter is proportional to the integral of
the one-particle density distributions of the colliding systems,
and the integration is performed along a straight-line trajectory
of motion. Generally, the integral contains also the form factor
of the NN scattering amplitude and thus its form is akin to that
of the standard folding potential with the NN potential instead of
the NN amplitude. The NN amplitude itself is known from the
experimental data and therefore, the usage of a profile function
offers certain advantages over approaches based on the folding
potential. So, in nuclear physics, the HEA amplitude is applied to
energies larger than hundred MeV/N (see, e.g.
\cite{Alkhazov2002,Casten2000,Alkhazov97}. However, in the last
two decades the HEA was generalized and applied to lower energies.
The prescription to calculate the profile function consists in a
replacement of the straight-line trajectory impact parameter $b$
by the distance of closest approach $r_{c}$ in the Coulomb field
or by the respective distance $r_{cn}$ in the presence of the
nuclear field (real part of OP). Doing so a reasonable agreement
with the experimental data on the proton- and nucleus-nucleus
reaction cross sections has been obtained in the region of
energies from 10 to 1000 MeV/N (see, e.g.
\cite{Lukyanov2004,Shukla2003,DeVries,Vitturi87,Charagi92,Charagi97,
Vismes2000,Brink81}. However, this approach becomes fairly rough
when one calculates differential cross sections and also the total
cross sections at comparably low energies. Besides, in the case of
the microscopic OP given in a form of tables, this approach needs
a numerical solution of the classical equation of motion to get
the corresponding trajectory of motion. For this reason the method
is not efficient for applications because of quite complicated
calculations. In this case the better way is to explore the
equivalent HEA optical potential outlined in
\cite{Lukyanov2004,Shukla2003} and then to solve the respective
Schr\"odinger equation numerically using a standard code which
enables one to get the exact scattering amplitude and the total
reaction cross sections including the interference terms as well.

We used this approach in \cite{Lukyanov2007} to get the
microscopic HEA imaginary part of the OP (ImOP) and added the real
part of OP (ReOP) \cite{Satchler79,Khoa93}. The ReOP includes the
direct term and the exchange one which involves non-linearity
effects. Also, the role of the spin-orbit interaction has been
considered. Additionally, the density dependence of the effective
NN interaction, as well as the sensitivity of the results to the
predictions of different theoretical models for the density of
$^{6}$He have been studied. It was shown that the more
sophisticated Large-Scale Shell Model (LSSM)
\cite{Kara2000,Amos2000b} density of $^{6}$He is the most
preferable one because it has led to a better agreement with the
data. It was concluded in \cite{Lukyanov2007} that the use of the
microscopic folding ReOP ($V^{F}$) and the HEA ImOP ($W^{H}$) has
led to agreement with the data on $^{6}$He+p elastic scattering
cross sections for 41.6 and 71 MeV/N. However, the data at lowest
energy 25.2 MeV/N have been explained only on a qualitative level
which is related to the limitations of using the HEA ImOP for
energies around and less than 25 MeV/N. This has led to the
necessity to reduce strongly the depth of HEA ImOP. It was shown
in \cite{Lukyanov2007} that the OP in the form
$U_{opt}=N_{R}V^{F}+i N_{I}W^{H}$ with both $V^{F}$ and $W^{H}$
calculated microscopically and using only two free parameters
$N_{R}$ and $N_{I}$ which renormalize the ReOP and ImOP depths can
be reasonably applied to calculations of scattering cross sections
at energies $E_{inc}<100$ MeV/N, such as 41.6 and 71 MeV/N.

In the present work we apply the developed approach to study the
existing experimental data on $^{8}$He+p elastic scattering cross
sections at incident energies less than 100 MeV/N. Various model
densities of $^{8}$He, such as those obtained within the approach
of Tanihata {\it et al.} \cite{Tani92}, LSSM
\cite{Kara2000,Amos2000b} and the Jastrow correlation method (JCM)
\cite{Massen99,Moust2000} are used to calculate the OP's. We study
the role of the spin-orbit terms and in addition to our previous
study \cite{Lukyanov2007}, we consider two more parameters
$N_{R}^{SO}$ and $N_{I}^{SO}$ (when necessary) which renormalize
the depths of the real and imaginary parts of the SO potential,
respectively. In addition, the nuclear surface effects are also
studied by introducing an additional surface term in OP. This is
related to investigations of the lowest energy limit of the
applicability of the HEA OP in $^{8}$He+p elastic scattering. Also
we pay attention to the energy dependence of the parameters
$N_{R}$ and $N_{I}$ as well as to the respective volume integrals.
We note the necessity to analyze the differential cross sections
estimating simultaneously the values of the total reaction cross
section. This would give an additional test of the various
ingredients of the approach.

The theoretical scheme to calculate microscopically the real and
imaginary parts of the OP, as well as the spin-orbit term is given
in Section \ref{s:theory}. The results of the calculations of OP's
and elastic scattering differential cross sections, including
those from some methodical ones, and their discussion are given in
Sec. \ref{s:results}. The summary of the work and conclusions of
the results are presented in Sec. \ref{s:concl}.

\section{THEORETICAL SCHEME\label{s:theory}}

\subsection{Direct and exchange parts of the real OP (ReOP)}
Here we give briefly the main expressions for the real part of the
nucleon-nucleus OP that is assumed to be a result of a single
folding of the effective NN potential and the nuclear densities.
It involves the direct and exchange parts (for more details, see,
e.g. \cite{Satchler79,Khoa93,Khoa2000} and also
\cite{Lukyanov2007}):
\begin{equation}
V^F(r)=V^D(r)+V^{EX}(r).
\label{eq:new}
\end{equation}
In Eq.~(\ref{eq:new}) the direct part ($V^{D}$) is composed of the
isoscalar (IS) and isovector (IV) contributrions, correspondingly:
\begin{equation}
V^D_{IS}(r)=\int \rho_2({\bf r}_2)g(E)F(\rho_2)v_{00}^D(s)d{\bf
r_{2}},
\label{eq:1}
\end{equation}
\begin{equation}
V^D_{IV}(r)=\int \delta\rho_2({\bf
r}_2)g(E)F(\rho_2)v_{01}^D(s)d{\bf r_{2}},
\label{eq:2}
\end{equation}
where ${\bf s}={\bf r}+{\bf r}_2$,
\begin{equation}
\rho_2({\bf r}_2)=\rho_{2,p}({\bf r}_{2,p})+\rho_{2,n}({\bf
r}_{2,n}),
\label{eq:3}
\end{equation}
\begin{equation}
\delta\rho_2({\bf r}_2)=\rho_{2,p}({\bf r}_{2,p})-\rho_{2,n}({\bf
r}_{2,n}).
\label{eq:4}
\end{equation}
In Eqs.~(\ref{eq:3}) and (\ref{eq:4}) $\rho_{2,p}({\bf r}_{2,p})$
and $\rho_{2,n}({\bf r}_{2,n})$ are the proton and neutron
densities of the target nucleus. In Eqs.~(\ref{eq:1}) and
(\ref{eq:2}) the energy dependence of the effective NN interaction
is taken in the usually used form:
\begin{equation}
g(E)=1-0.003E.
\label{eq:5}
\end{equation}
Also, for the NN potentials $v_{00}^{D}$ and $v_{01}^{D}$ we use
the expression from \cite{Khoa2000} for the CDM3Y6-type of the
effective interaction based on the solution of the equation for
the $g$-matrix, in which the Paris NN potential has been used. The
density dependence of the effective interaction is taken in the
following form:
\begin{equation}
F(\rho)=C\left [1+\alpha e^{-\beta\rho({\bf r})}-\gamma\rho({\bf
r})\right ],
\label{eq:6}
\end{equation}
where $C$=0.2658, $\alpha$=3.8033, $\beta$=1.4099 fm$^3$, and
$\gamma$=4.0 fm$^3$.

The isoscalar part of the exchange contribution to the ReOP has
the form (see, e.g. \cite{Lukyanov2007}):
\begin{eqnarray}
V^{EX}_{IS}(r)&=&g(E)\int \rho_2({\bf r}_2,{\bf r}_{2}-{\bf s})
F\Bigl[\rho_2\Bigl({\bf r}_2-\frac{{\bf s}}{2}\Bigr )\Bigr ]
\nonumber
\\ & \times & v_{00}^{EX}(s) j_{0}[k(r)s] d{\bf r}_{2},
\label{eq:7}
\end{eqnarray}
where the density matrix $\rho_2({\bf r}_2,{\bf r}_{2}-{\bf s})$
is usually approximated by the expression:
\begin{equation}
\rho_2\bigl({\bf r}_2, {\bf r}_2-{\bf s}\bigr)~\simeq~
\rho_2\Bigl(\Bigl|{\bf r}_2-\frac{{\bf s}}{2}\Bigr|\Bigr) {\hat
j}_1\Bigl[k_{F,2}\Bigl(\Bigl|{\bf r}_2-\frac{{\bf
s}}{2}\Bigr|\Bigr)\cdot s\Bigr]
\label{eq:8}
\end{equation}
with
\begin{equation}
{\hat j}_1(x)={3\over x}j_1(x)={3\over x^3}(\sin x-x\cos x)
\label{eq:9}
\end{equation}
and $v_{00}^{EX}(s)$ is the isoscalar part of the exchange
contribution to the effective NN interaction. The local momentum
$k(r)$ of the incident nucleon in the field of the Coulomb and
nuclear potential (ReOP) is \cite{Khoa2002}:
\begin{equation}
k^2(r)=\left(\frac{2m}{\hbar^2}\right)\left[E_{c.m.}-V_{c}(r)-V(r)\right
]\left({1+A_2\over A_2}\right ). \label{eq:11}
\end{equation}
Substituting Eq.~(\ref{eq:11}) in Eq.~(\ref{eq:7}) the iteration
procedure was used to get the final result for the folding
potential. One can see that in this procedure the required
microscopic potential $V(r)$ (that has to be calculated according
to Eq.~(\ref{eq:new})) appears in the expression for $k^{2}(r)$
[Eq.~(\ref{eq:11})] and, correspondingly, in the integrand of the
integral in Eq.~(\ref{eq:7}), i.e. in the expression for the
exchange contribution to the OP. Thus, non-linearity effects occur
as typical ingredients of the model and they have to be taken
carefully into account. In our consideration, for the highest
energy 73 MeV/N eight iterations and for the lowest one 15.7 MeV/N
thirteen iterations were large enough in the calculations of the
folding potentials.

In Eq.~(\ref{eq:8}) $k_{F,2}$ is the average relative momentum of
a nucleon in a nucleus \cite{Campy78,Khoa2002}:
\begin{equation}
k_{F,2}(r)=\left\{{5\over 3\rho}\left[\tau(\rho)- {1\over
4}{\bf\nabla}^2\rho(r)\right ]\right \}^{1/2},
\label{eq:12}
\end{equation}
where we choose for the kinetic energy density $\tau(\rho)$ the
expression from the extended Thomas-Fermi approximation
\cite{Ring80,Khoa2001}:
\begin{eqnarray}
{\tau(\rho)\over 2}&\simeq&\tau_q(\rho_q)={3\over
5}\left(3\pi^2\right )^{2/3} \left[\rho_q(r)\right
]^{5/3}\nonumber \\
&+& {|{\bf\nabla}\rho_q(r)|^2\over 36\rho_q(r)}+
{{\bf\nabla}^2\rho_q(r)\over 3}
\label{eq:13}
\end{eqnarray}
valid for each kind of particles $q=n,p$. It is shown in
\cite{Lukyanov2007} how the isovector part of the exchange ReOP
can be obtained.

\subsection{Density distributions of $^{8}$He}

In the calculations of the OP's we use the following point-nucleon
density distributions of $^{8}$He:

i) the Tanihata densities deduced in \cite{Tani92} by means of
comparison of the measured total reaction cross section of
$^{6,8}$He+$^{12}$C at 800A MeV with the respective expression
from \cite{Karol75} derived in the framework of the optical limit
of the Glauber theory:
\begin{eqnarray}
\rho^X_{point}&=&{2\over \pi^{3/2}}\biggl\{{1\over a^3}\exp
\left[-\left({r\over a}\right )^2\right ]\nonumber \\
&+&{1\over b^3}{(X-2)\over 3}\left({r\over b}\right )^2
\exp\left[-\left({r\over b}\right )^2\right ]\biggr\}.
\label{eq:14}
\end{eqnarray}
Here $X=Z,N$ and the parameter values of $a$ and $b$ can be
determined from
\begin{equation}
a^2={a^\ast}^2\left(1-{1\over A}\right ), \qquad
b^2={b^\ast}^2\left(1-{1\over A}\right ),
\label{eq:15}
\end{equation}
where ${a^\ast}$=1.53 fm and ${b^\ast}$=2.06 fm; hence $a$=1.43 fm
and $b$=1.93 fm for $^{8}$He. So, the proton distribution is
defined by the first term only, while an excess of neutrons is
described by the additional second term. The rms radii of the
point-proton and point-neutron densities of $^{8}$He are equal to
1.76 fm and 2.69 fm, correspondingly;

ii) the LSSM densities calculated in a complex 4$\hbar\omega$
shell model space \cite{Kara2000,Amos2000b} using the Woods-Saxon
(WS) basis of single-particle wave functions with realistic
exponential asymptotic behavior;

iii) the densities obtained in \cite{Massen99,Moust2000} with
accounting for the NN central-type short-range Jastrow
correlations.

\subsection{Optical potential within the high-energy
approximation}

In Ref.~\cite{Lukyanov2007} the so-called complex HEA optical
potential has been applied to explain the available data on the
$^{6}$He+p elastic differential  cross sections and energies less
than 100 MeV/N. The HEA OP was derived in \cite{Lukyanov2004} on
the basis of the eikonal phase inherent in the optical limit of
the Glauber theory. Then, by means of this potential or taking
only its imaginary part together with the folding real part of OP,
the cross sections were calculated using the code DWUCK4
\cite{Kunz93} for solving the Schr\"{o}dinger equation. Thus, we
don't apply the Glauber theory for calculating the scattering
amplitude at relatively low energies but utilize the equivalent
HEA OP to solve numerically the respective wave equation. In this
case, the use of the ordinary Glauber theory leads to insuperable
problems in performing integration in the eikonal phase mentioned
in the Introduction. Indeed, there one should take into account
the distortion of the integration path along classical
trajectories in the field of the Coulomb and nuclear potentials
(see, e.g.
\cite{Lukyanov2004,DeVries,Vitturi87,Charagi92,Charagi97,Vismes2000,
Brink81}. At the same time, to calculate the HEA OP one can use
the definition of the eikonal phase as an integral of the
nucleon-nucleus potential over the trajectory of the straight-line
propagation, and have to compare it with the corresponding Glauber
expression for the phase in the optical limit approximation. Doing
so, the HEA OP can be obtained as a folding of form factors of the
nuclear density and the NN amplitude $f_{NN}(q)$
\cite{Lukyanov2004,Shukla2003}:
\begin{eqnarray}
U^H_{opt}&=&V^H+iW^H=-{\hbar v\over
(2\pi)^2}(\bar\alpha_{NN}+i)\bar\sigma_{NN}\nonumber \\
& \times & \int_0^\infty dq q^2 j_0(qr) \rho_2(q) f_{NN}(q).
\label{eq:16}
\end{eqnarray}
In (\ref{eq:16}) $\bar\sigma_{NN}$ and $\bar\alpha_{NN}$ are,
respectively, the NN total scattering cross section and the ratio
of the real to imaginary parts of the forward NN scattering
amplitude both averaged over the isospin of the nucleus. They both
have been parametrized in \cite{Charagi92,Shukla2001} as functions
of energies in a wide range from 10 MeV to 1 GeV and also at
energies lower than 10 MeV. The values of these quantities can
also account for the in-medium effect by a factor from
\cite{Xiangzhow98}.

\subsection{The spin-orbit term}

Following Refs.~\cite{Kunz93,Becchetti69,Koning2003} the expression
for the spin-orbit contribution to the OP can be written in the
form:
\begin{equation}
V_{LS}(r)=2\lambda_{\pi}^{2}\left[V_{0}\frac{1}{r}\frac{df_{R}(r)}{dr}+iW_{0}
\frac{1}{r}\frac{df_{I}(r)}{dr}\right]({\bf l}\cdot{\bf s}),
\label{eq:17}
\end{equation}
where $\lambda_{\pi}^{2}$=2 fm$^{2}$ is the squared pion Compton
wavelength, $V_{0}$ and $W_{0}$ are the real and imaginary parts
of the microscopic OP at $r$=0, and $f(r)$ is the form of the real
[$f_{R}(r)$] and imaginary [$f_{I}(r)$] parts of the microscopic
OP taken as WS forms $f(r,R_{R},a_{R})$ and $f(r,R_{I},a_{I})$. In
our calculations the parameters (half-radius $R_{R}(R_{I})$ and
diffuseness $a_{R}(a_{I})$) are obtained by fitting the WS
potential to the microscopically calculated real and imaginary
contributions to the OP $V(r)$ and $W(r)$.

\section{RESULTS AND DISCUSSION\label{s:results}}

\begin{figure}
\includegraphics[width=1.0\linewidth]{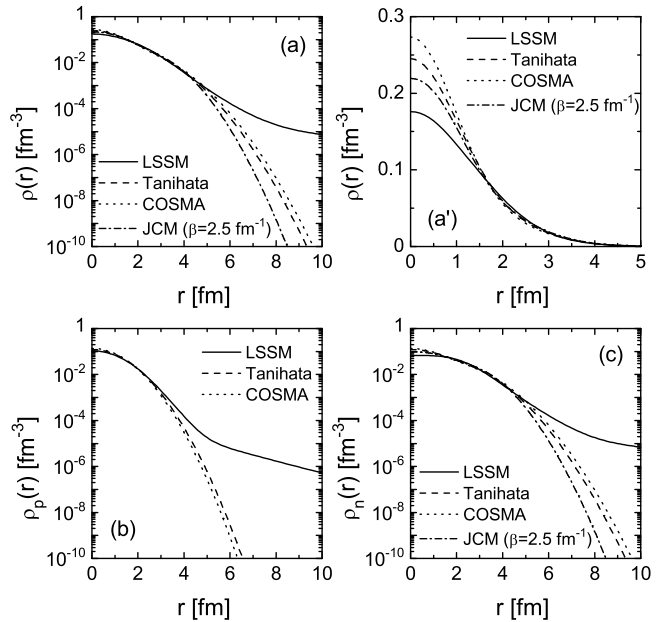}
\caption{Total ((a) and (a$^\prime$)), point-proton (b) and
point-neutron (c) densities of $^8$He from the model of Tanihata
\cite{Tani92}, COSMA \cite{Zhukov93,Zhukov94}, LSSM
\cite{Kara2000,Amos2000b} and JCM calculations
\cite{Massen99,Moust2000}.}
\label{fig1_dens}
\end{figure}

\begin{figure}
\includegraphics[width=1.0\linewidth]{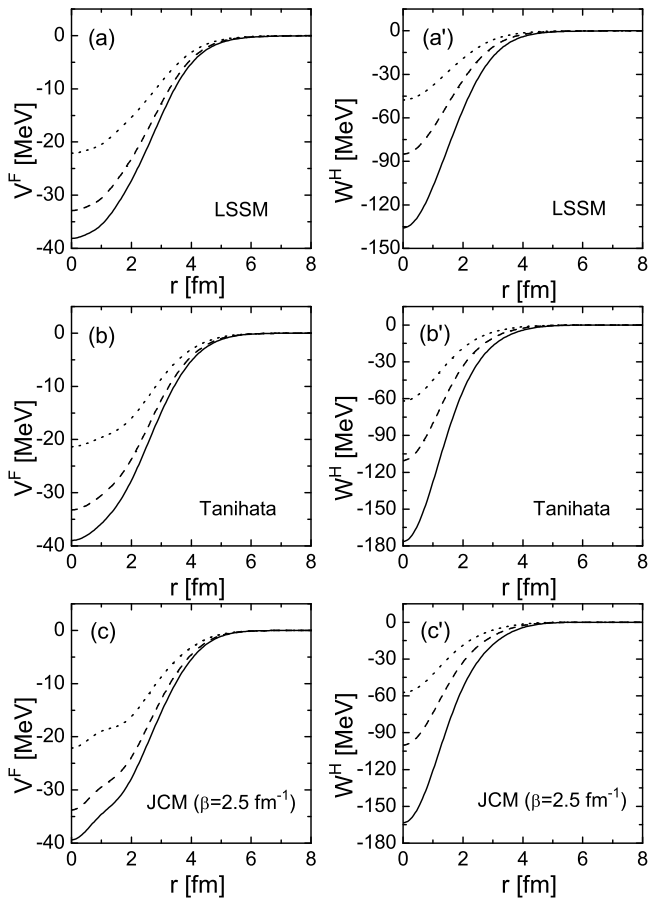}
\caption{Microscopic real part ($V^F$) of OP ((a), (b) and (c))
and HEA imaginary part ($W^H$) ((a$^\prime$), (b$^\prime$) and
(c$^\prime$)) calculated using the LSSM, Tanihata and JCM
($\beta=2.5$ fm$^{-1}$) densities of $^8$He for energies $E$=15.7
(solid lines), 32 (dashed lines) and 73 MeV/N (dotted lines).
\label{fig2_pot} }
\end{figure}

In this Section we present the results of the calculations of the
microscopic OP's and the respective $^{8}$He+p elastic scattering
differential cross sections at energies $E_{inc}<100$ MeV/N. In
principle, the OP's do not contain free parameters, but they
depend on the density distribution of the target nucleus. This
allows one to test advanced theoretical methods that give
predictions for the density distribution. In the case of $^8$He we
used the semi-empirical model of Tanihata \cite{Tani92}, the
large-scale shell model \cite{Kara2000,Amos2000b}, as well as the
results  of the approach \cite{Massen99,Moust2000} within the JCM.
In Fig.~\ref{fig1_dens} in logarithmic and linear scales are shown
the proton $\rho_{p}(r)$, neutron $\rho_{n}(r)$ and matter
$\rho(r)$ densities of $^8$He obtained in different models. Also,
for comparison, the known COSMA densities \cite{Zhukov93,Zhukov94}
are presented. We note that among them only the LSSM densities
have a realistic exponential asymptotics, whereas the others have
a Gaussian one. The results for the JCM densities are given for
the value of the correlation parameter $\beta$=2.5 fm$^{-1}$ in
the Jastrow correlation factor $1-e^{-\beta^{2}r^{2}}$, where $r$
is the distance between neutrons. It was shown in
Refs.~\cite{Massen99,Moust2000} that the inclusion of this factor
causes a slight increase of the density in the central part of the
nucleus. Simultaneously, as can be seen in Fig.~\ref{fig2_pot},
this leads to a small decrease of the depth of the imaginary part
of OP in comparison with the case of the Tanihata density. In the
same Figure we show as examples the real $V^F$ and imaginary $W^H$
parts of the $^8$He+p OP's for energies 15.7, 32 and 73 MeV/N
calculated using different densities. $V^F$ is calculated by a
folding procedure and $W^H$ within the HEA (see Section
\ref{s:theory}). It is seen that the increase of the energy leads
to reduced depths and slopes of ReOP and ImOP.

We calculated the $^8$He+p elastic scattering differential cross
sections utilizing the program DWUCK4 \cite{Kunz93} and using the
microscopically obtained real $V^{F}$ and imaginary $W^{H}$
contributions to the optical potential:
%
%
\begin{eqnarray} \label{uopt}
U_{opt}(r) &=& N_R V^F(r) + i N_I W^H(r) +  \nonumber \\
&+& 2\lambda_{\pi}^{2} \left\{ N_R^{SO} V^F_0 \frac 1 r \frac
{df_R(r)} {dr} \right. \\
&+& \left. i N_I^{SO} W^H_0 \frac 1 r \frac {df_I(r)} {dr}\right
\} ({\bf l.s}), \nonumber
\end{eqnarray}
where $V_{0}^{F}$ and $W_{0}^{H}$ are the depths of the SO optical
potential obtained simultaneously with $R_{R}(R_{I})$ and
$a_{R}(a_{I})$ from the approximation of the volume real and
imaginary OP's by Woods-Saxon form.

So, further in the present work we consider the set of the $N$
coefficients as parameters to be found out from comparisons with
the experimental data. We consider such a model as the appropriate
physical basis, which constraints the fitting procedure by the
established model forms of searching potentials. We would like to
emphasize here that in our work we do not aim a perfect agreement
with the empirical data. In this sense, the introduction of the
fitting parameters ($N$'s) related to the depths of the different
components of the OP's can be considered as a way to introduce a
quantitative measure of the deviations of the predictions of our
approach from the reality (e.g. the differences of $N$'s from
unity for given energies, as can be seen below).

\begin{figure}
\includegraphics[width=1.0\linewidth]{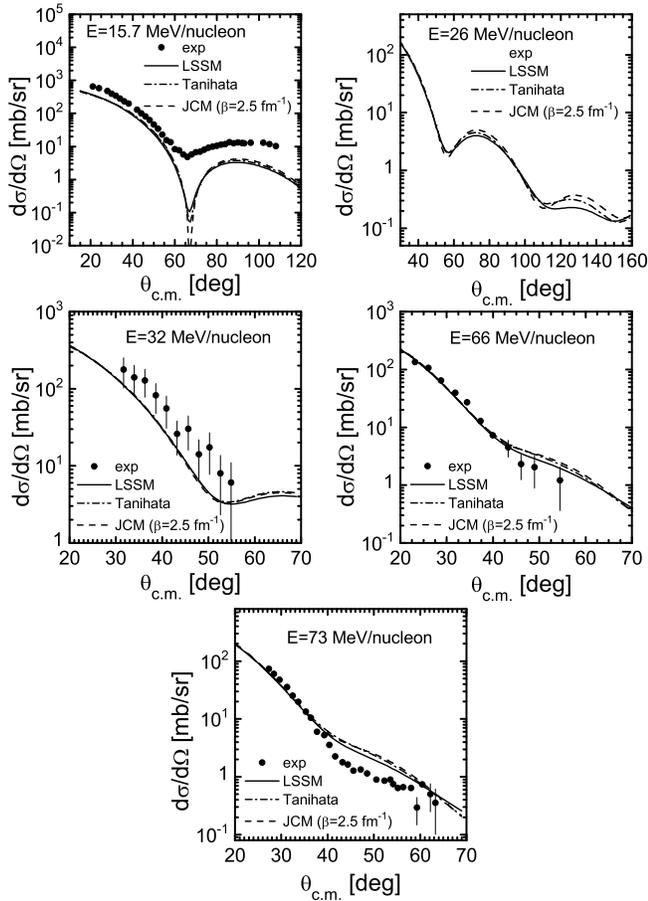}
\caption{The $^8$He+p elastic scattering cross sections at
different energies calculated using $U_{opt}$ [Eq.~(\ref{uopt})]
for values of the parameters $N_R$=$N_I$=1 and
$N_R^{SO}$=$N_I^{SO}$=0. The used densities of $^8$He are LSSM
(solid line), Tanihata (dash-dotted line) and JCM ($\beta$=2.5
fm$^{-1}$) (dashed line). Experimental data are taken for 15.7
\cite{Skaza2005}, 26 \cite{Ter2000}, 32 \cite{Korsh97a,Korsh97b},
66 \cite{Korsh97a,Korsh97b} and 73 MeV/N
\cite{Korsh97a,Korsh97b,Korsh93}. \label{fig3} }
\end{figure}

The discussion that follows is based on the fitting procedure,
where the additionally introduced strength parameters $N_R$,
$N_I$, $N_R^{SO}$, $N_I^{SO}$ are varied step by step. So, we
start from the case $N_R$=$N_I$=1, $N_R^{SO}$=$N_I^{SO}$=0, then
fit successively both coefficients $N_R$ and $N_I$, and after that
the values of $N_R^{SO}$ and $N_I^{SO}$. First, we give in
Fig.~\ref{fig3} the results of our methodical calculations of the
cross sections for different energies (15.7, 26, 32, 66 and 73
MeV/N) using the densities of $^8$He from LSSM, Tanihata and JCM
approaches in the case when $N_R$=$N_I$=1 and
$N_R^{SO}$=$N_I^{SO}$=0 (i.e. without spin-orbit interaction). It
can be seen that the behavior of the cross sections for a given
energy and interval of angles is weakly sensitive to the choice of
the model for the density of $^8$He. In spite of this uncertainty
we choose for the further applications the LSSM density since it
has a realistic exponential behavior in the peripheral region of
the nucleus.

\begin{figure}
\includegraphics[width=1.0\linewidth]{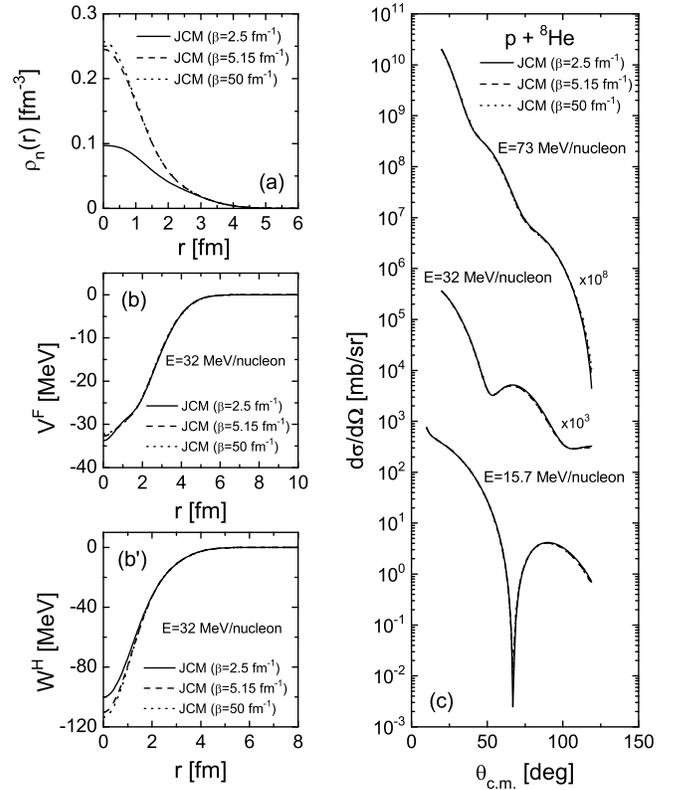}
\caption{Point-neutron density of $^8$He (a), $V^F$ and $W^H$ of
OP ((b) and (b$^\prime$)) for energy $E$=32 MeV/N and $^8$He+p
elastic scattering cross sections (c) at energies $E$=15.7, 32 and
73 MeV/N calculated using JCM densities of $^8$He for three values
of the correlation parameter $\beta$. \label{fig4} }
\end{figure}

The second methodical study is a test of the effect of Jastrow
central short-range NN correlations on mechanism of the considered
process of scattering. As known, the main parameter that governs
the contribution of these correlations is $\beta$, and we change
it in wide limits from 2.5 fm$^{-1}$ to 50 fm$^{-1}$. It is seen
in Fig.~\ref{fig4} that these changes result in an increase of the
neutron density of about 2.5 times in the central part of $^{8}$He
but this has no important effect on the calculated OP's and on the
shape of the respective differential cross sections. Therefore, in
the further calculations we do not account for the short-range
correlation effects.

\begin{figure}
\includegraphics[width=1.0\linewidth]{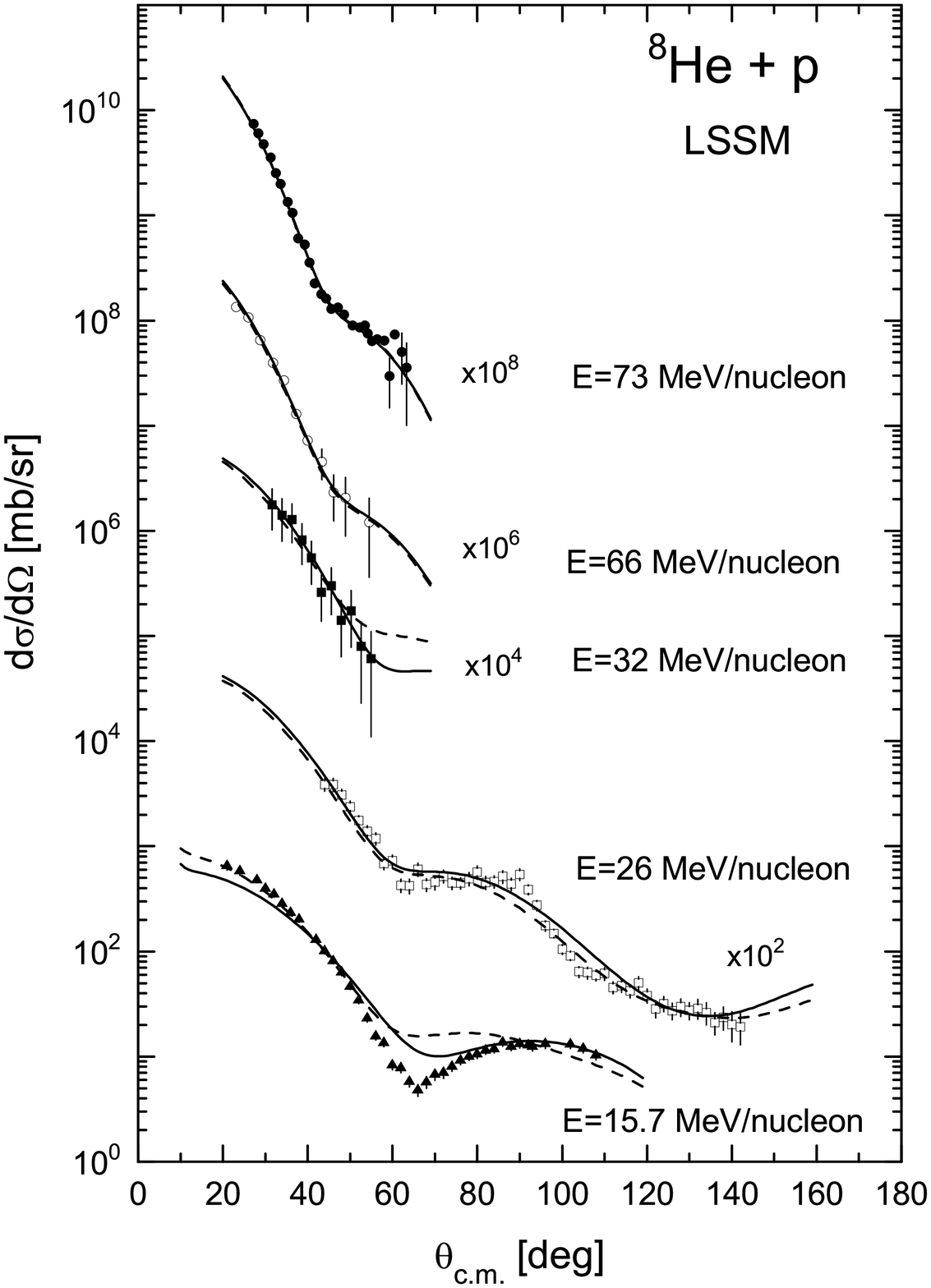}
\caption{The $^8$He+p elastic scattering cross sections at
different energies calculated using $U_{opt}$ [Eq.~(\ref{uopt})]
for various values of the renormalization parameters $N_R$, $N_I$,
$N_R^{SO}$ and $N_I^{SO}$ (presented in Table \ref{tab1}) giving
the best agreement with the data. The used density of $^8$He is
LSSM. Experimental data are taken for 15.7 \cite{Skaza2005}, 26
\cite{Ter2000}, 32 \cite{Korsh97a,Korsh97b}, 66
\cite{Korsh97a,Korsh97b} and 73 MeV/N
\cite{Korsh97a,Korsh97b,Korsh93}. \label{fig5} }
\end{figure}

Later, as a next step, we allow the "depth" of each of the parts
of the OP (\ref{uopt}) in our semi-microscopic models to vary in
order to find the optimal values of the parameters $N_R$, $N_I$,
$N_R^{SO}$ and $N_I^{SO}$ by a fitting procedure to the available
experimental data for the cross sections. In Fig.~\ref{fig5} we
present the results of our calculations of $^8$He+p elastic
scattering cross sections for various energies and the LSSM
density with the fitted values of the parameters $N_R$, $N_I$,
$N_R^{SO}$ and $N_I^{SO}$. The values of these renormalization
parameters are given in Table~\ref{tab1} together with the
predicted total reaction cross sections. The results obtained
using the values of the parameters from the first line of this
Table for each energy are given by solid line in Fig.~\ref{fig5},
while those from the second line for each energy are given by
dashed line.

As is known, however, the problem of the ambiguity of the values
of the parameters {\it N} arises when the fitting procedure is
applied to a limited number of experimental data. For instance, in
the case of the LSSM density, the values of $N_R$=1.0 and 0.9, and
correspondingly $N_I$=0.236 and 0.1 (with $N_R^{SO}$=0.107 and
$N_I^{SO}$=0.040) lead to similar results in the case of 15.7
MeV/N. For $E$=32 MeV/N the results are similar when $N_R$=1.0,
$N_I$=0.374 and $N_R$=0.438, $N_I$=0.036; for $E$=66 MeV/N the
results are similar when $N_R$=0.876, $N_I$=0.071 and $N_R$=0.854,
$N_I$=0.086; for $E$=73 MeV/N they are similar when $N_R$=0.875,
$N_I$=0.020; $N_R$=0.869, $N_I$=0.010 (with $N_R^{SO}$=0.009 and
$N_I^{SO}$=0.002). Our calculations produce similar results when
using the Tanihata density. We note that in some cases it has been
enough to vary only the volume part of the OP, i.e. the values of
the parameters $N_R$ and $N_I$ without the spin-orbit parts of the
OP. When all four parameters $N$ are fitted the results for a
given energy are similar, as already mentioned above. Thus, the
problem to choose the most physical values of the parameters $N$
arises. It is known that because the procedure of fitting belongs
to the class of the ill-posed problems (see, e.g.
Ref.~\cite{Tikhonov77}), it is necessary to impose some physical
constraints on the choice of the set of parameters $N$. One of
them is the total cross section of scattering and reaction.
However, the corresponding values are missing at the energy
interval considered in our work. To our knowledge, the total
reaction cross section $\sigma_R$ of $^8$He+p process is known
only at energy 670 MeV and it is about 200 mb \cite{Alkhazov2002}.

\begin{table}
\caption{The renormalization parameters $N_R$, $N_I$, $N_R^{SO}$
and $N_I^{SO}$ obtained by fitting the experimental data in
Fig.~\ref{fig5} in the case of LSSM density. The energies are in
MeV/N and the total reaction cross sections $\sigma_{R}$ are in
mb.}
\label{tab1}
\begin{center}
\begin{tabular}{cccccccccccc}
\hline\noalign{\smallskip}
$E$   & & & $N_R$  & & $N_I$  & & $N_R^{SO}$ & & $N_I^{SO}$ & &
$\sigma_{R}$ \\
\noalign{\smallskip}\hline\noalign{\smallskip}
15.7  & & &  1.0   & & 0.236  & & 0      &  &  0      &  &  603.6
\\
15.7  & & &  0.9   & & 0.1    & & 0.107  &  &  0.040  &  &  693
\\
26    & & &  0.422 & & 0.104  & & 0.090  &  &  0.010  &  &  275.11
\\
26    & & &  0.439 & & 0.144  & & 0.087  &  &  0.023  &  &  377.22
\\
32    & & &  0.438 & & 0.036  & & 0.096  &  &  0      &  &  71.9
\\
32    & & &  1.0   & & 0.374  & & 0      &  &  0      &  &  419.5
\\
66    & & &  0.876 & & 0.071  & & 0      &  &  0      &  &  55.7
\\
66    & & &  0.854 & & 0.086  & & 0      &  &  0      &  &  65.9
\\
73    & & &  0.875 & & 0.02   & & 0      &  &  0      &  &  1.48
\\
73    & & &  0.869 & & 0.01   & & 0.010  &  &  0.002  &  &  1.22
\\
\noalign{\smallskip}\hline
\end{tabular}
\end{center}
\end{table}

Another physical criterion that has to be imposed on the choice of
the values of the parameters $N$ is the behavior of the volume
integrals \cite{Satchler79}
%
\begin{eqnarray}\label{volint}
J_V&=&\frac{4\pi}{A}\int dr r^2 [N_{R}V^{F}(r)], \\
J_W&=&\frac{4\pi}{A}\int dr r^2 [N_{I}W^{H}(r)]
\end{eqnarray}
as functions of the energy.

It has been pointed out (see, e.g. Romanovsky {\it et al.}
\cite{Romanovsky} and references therein) that the values of the
volume integral $J_V$ decrease with the increase of the energy in
the interval $0<E<100$ MeV/N, while $J_W$ is almost constant in
the same interval. Imposing this behavior of $J_V$ and $J_W$ on
our OP's (i.e. on their ``depth'' parameters $N_R$ and $N_I$), we
obtain by the fitting procedure the values of the parameters given
in Table \ref{tab2}.

\begin{table}
\caption{The parameters $N_R$, $N_I$, $N_R^{SO}$ and $N_I^{SO}$,
the volume integrals $J_{V}$ and $J_{W}$ (in MeV.fm$^{3}$) as
functions of the energy $E$ (in MeV/N) (selected in correspondence
to the behavior shown in Fig.~\ref{fig6}(b,c)), and the total
reaction cross sections $\sigma_{R}$ (in mb) for the $^8$He+p
scattering in the case of LSSM density (the results of the fit are
shown in Fig.~\ref{fig6}(a)).} \label{tab2}
\begin{center}
\begin{tabular}{cccccccccc}
\hline\noalign{\smallskip}
$E$   & & $N_R$ & $N_I$ & $N_R^{SO}$ & $N_I^{SO}$ & &  $J_{V}$ &
$J_{W}$ & $\sigma_{R}$ \\
\noalign{\smallskip}\hline\noalign{\smallskip}
15.7  &  & 0.630   &  0.064  &  0.139  &  0.070  & & 411.1  &  58.6
& 722.0  \\
15.7  &  & 0.630   &  0.052  &  0.166  &  0.057  & & 411.1  &  47.6
& 701.2  \\
26    &  & 0.644   &  0.128  &  0.035  &  0.026  & & 377.7  &  84.35
& 381.2  \\
32    &  & 0.648   &  0.120  &  0.062  &  0.022  & & 358.3  &  69
& 302.7  \\
66    &  & 0.852   &  0.131  &  0      &  0      & & 344.2  &  45
& 95.2   \\
73    &  & 0.869   &  0.090  &  0.004  &  0      & & 330.0  &  29
& 60.9   \\
73    &  & 0.869   &  0.063  &  0.010  &  0      & & 330.0  &  20.25
& 43.9   \\
\noalign{\smallskip}\hline
\end{tabular}
\end{center}
\end{table}

\begin{figure}
\includegraphics[width=1.0\linewidth]{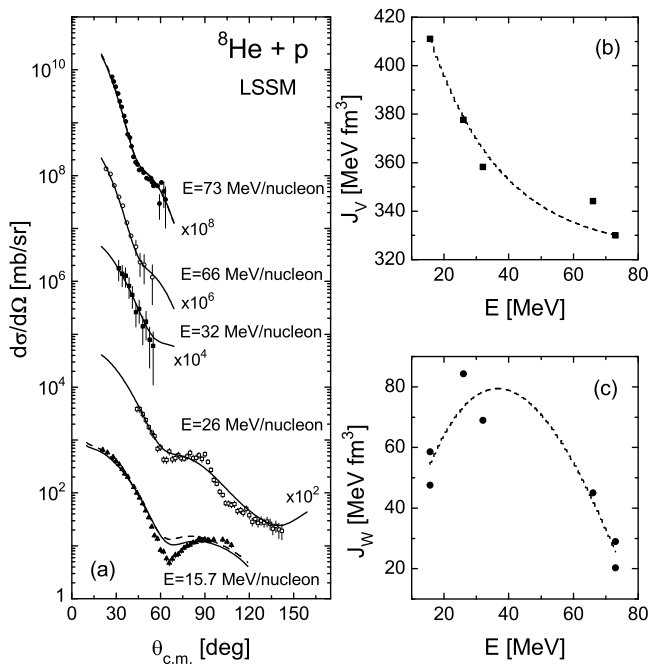}
\caption{The $^8$He+p elastic scattering cross sections (a) at
different energies using LSSM density of $^8$He and parameters
from Table~\ref{tab2}. Experimental data are taken for 15.7
\cite{Skaza2005}, 26 \cite{Ter2000}, 32 \cite{Korsh97a,Korsh97b},
66 \cite{Korsh97a,Korsh97b} and 73 MeV/N
\cite{Korsh97a,Korsh97b,Korsh93}. The obtained values of the
volume integrals $J_V$ (b) and $J_W$ (c) (given by points) are
shown as functions of the incident energy, while the dashed lines
give the trend of this dependence. \label{fig6} }
\end{figure}

The results of the calculations of the cross sections are
presented in Fig.~\ref{fig6} for the case of the LSSM density
together with the volume integrals $J_V$ and $J_W$ as functions of
the energy. The results obtained using the values of the
parameters from the first line of Table~\ref{tab2} for the
energies 15.7 and 73 MeV/N are given by solid line in
Fig.~\ref{fig6}(a), while those from the second line for these
energies are given by dashed line. In comparison to the data in
Table~\ref{tab1} one can see that the total reaction cross
sections decrease monotonically with the energy increased. Also,
we reach the smooth change of the values of the volume integrals
with the energy increase. Moreover, with the energy increase one
sees the monotonic increase of the renormalization coefficients
$N_{R}$ of the volume real part of OP together with an "average"
decreasing of $N_{I}$ inherent in the imaginary part of OP. Almost
a regular behavior is obtained for the spin-orbit correction
coefficients $N_{R}^{SO}$ and $N_{I}^{SO}$. So, the $N_{R}$
coefficient is going to 1 in coincidence with a general conception
of a folding procedure. But the obtained small values of $N_{I}$
and problems with the fitting of our OP at low energies deserves a
special attention.

It is known that the fitting of the phenomenological OP's to the
data of proton scattering on light nuclei leads to ``shallow''
imaginary parts of the OP's whose depths are sufficiently smaller
than that of the real part of the OP. This has been observed in
our previous works \cite{Lukyanov2007,izvran_he6} for the case of
$^6$He+p elastic scattering. This is the case also in our present
calculations. Another remark is connected with the difficulties in
the description of the cross sections at low energies. In this
case we cannot fit the data using only the volume form of OP.
Instead, if one adds the contribution of a surface part of OP,
then a better agreement with the data can be achieved. We would
like to remind that such an admixture had been used in the earlier
applications of the phenomenological OP (see, e.g.
\cite{Koning2003}).

For this reason we consider also the contribution of the surface
potential:
\begin{equation} \label{surface}
U^{\prime}_{opt}(r)=U_{opt}(r) - i 4a N_S \frac{dV^{F}(r)}{dr},
\end{equation}
where the first term in the right-hand side is the expression for
the OP given by Eq.~(\ref{uopt}) (in which the ImOP is taken in
the form of $V^{F}(r)$) and the second term is responsible for the
surface effects. We would like to note that, in particular, for
the lowest incident energy, the combination of the microscopically
folded real and imaginary parts in the form of $V^{F}$ is more
appropriate. In Eq.~(\ref{surface}) $a$ is the diffuseness
parameter of $V^{F}(r)$ fitted by WS form.

\begin{figure}[h]
\includegraphics[width=0.75\linewidth]{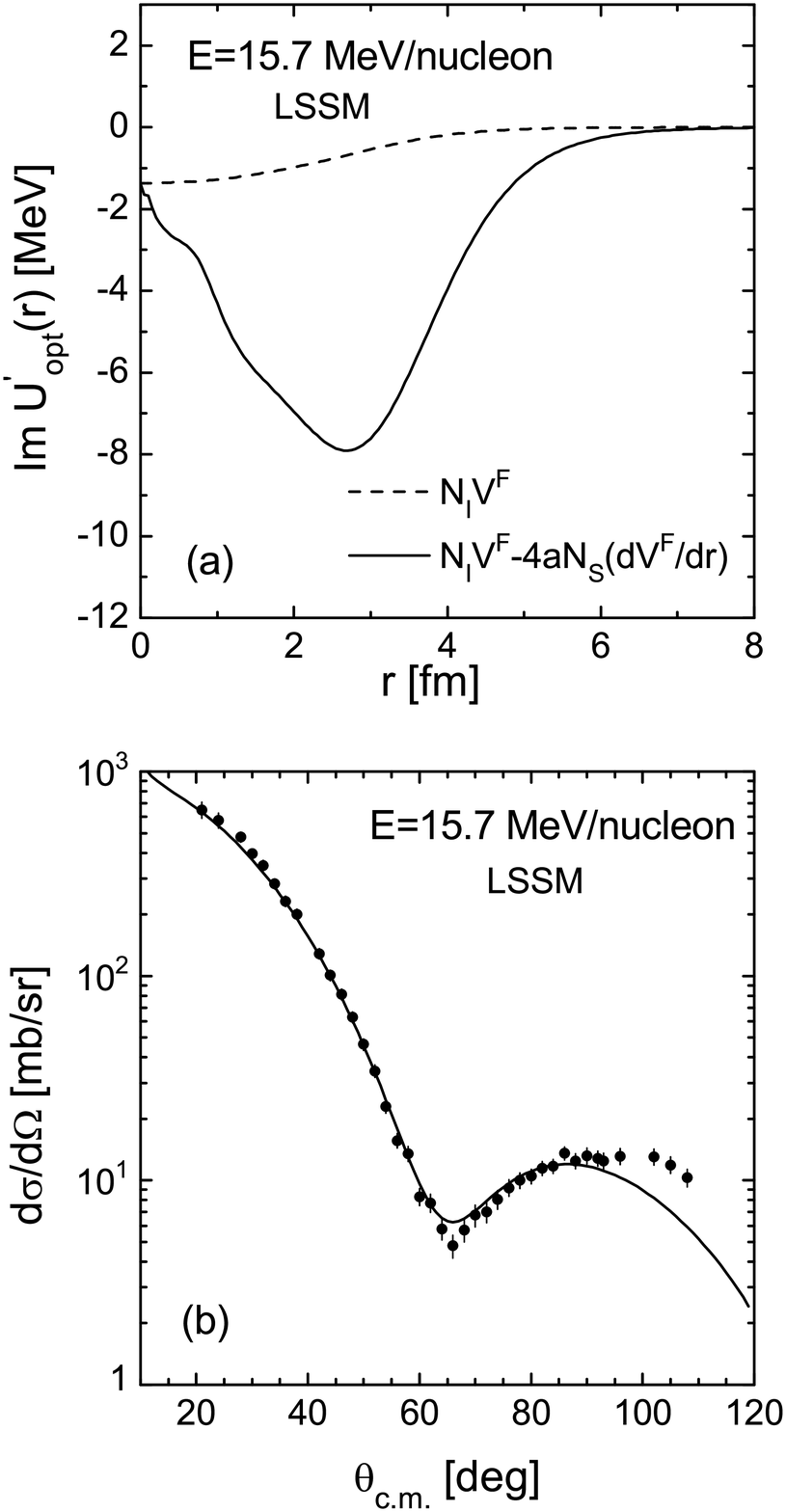}
\caption{(a) Volume (dashed line) and total (solid line) imaginary
parts of the OP $U^{\prime}_{opt}(r)$ [Eq.~(\ref{surface})]
calculated using the LSSM density of $^8$He for energy $E$=15.7
MeV/N; (b) The $^8$He+p elastic scattering cross section at energy
$E$=15.7 MeV/N using LSSM density of $^8$He with fitted values of
$N_R$, $N_I$, $N_R^{SO}$, $N_I^{SO}$ and $N_S$ given in the text.
Experimental data are taken from \cite{Skaza2005}. \label{fig7} }
\end{figure}

We present in Fig.~\ref{fig7}(a) our results for both the volume
and all components of the imaginary part of the potential
$U^{\prime}_{opt}(r)$ and in Fig.~\ref{fig7}(b) for the cross
section in the case of $E$=15.7 MeV/N obtained using the LSSM
density of $^8$He. The calculations are performed by fitting the
strength parameters $N_R$, $N_I$, $N_R^{SO}$, $N_I^{SO}$ entering
Eqs.~(\ref{uopt}) and (\ref{surface}) and the depth parameter
$N_S$ of the surface term of the OP [Eq.~(\ref{surface})]. In this
case $N_R$=1.078, $N_I$=0.036, $N_R^{SO}$=$N_I^{SO}$=0,
$N_S$=0.207, $a$=0.686 fm, $\sigma_{R}$=791.1 mb. It is seen from
Fig.~\ref{fig7} that the inclusion of the surface contribution to
the imaginary part of the OP improves the agreement with the
experimental data, especially for small angles and in the region
of the cross section minimum. Obviously, for more successful
description of the cross sections at low energies (15.7 and 26
MeV/N) our method has to be modified and improved by an inclusion
of virtual excitations of inelastic and decay channels of the
reactions.

From the results presented in this Section one can see that a
notable renormalization of the imaginary parts of the microscopic
OP and the necessity of its shape correction at lower energies are
needed for a reliable explanation of the data. In this connection
one should remind that both the folding and HEA potentials have
the same physical origin, namely, they are one-particle folding
potentials, and thus they do not account for more complicated
dynamical processes. We have already mentioned the role of the
inelastic and breakup channels. In the last years many works have
appeared where amplitudes of these processes were calculated
within the distorted wave approximation and also by using the
coupled channel methods. The latter provide a way of estimating
the elastic scattering cross sections, too (see, e.g.
\cite{Denikin2009} and references therein). On the other hand, if
one considers the elastic channel itself, the general and formally
established concept of the Feshbach theory \cite{Feshbach58} can
give us a basis for the following qualitative physical suggestion.
Indeed, in this theory the elastic scattering potential is
composed of two parts, the bare potential composed from
one-particle matrix elements and the so-called dynamical
polarization potential. Then, transforming this concept onto our
model of OP, one can suppose that the bare OP is the
microscopically calculated $V_{sp}=V^{F}+iW^{H}$ potential having
the strength $N_{R,I}\cong 1$. And the rest part $V_{pol}\cong
V_{fit}-V_{sp}$, being the difference between the fitted OP and
$V_{sp}$ may be identified with a polarization potential. In the
framework of this outline of the scattering mechanism one can
compare, for example, the imaginary part $W_{pol}$ with the
imaginary part $W_{breakup}$ obtained by fitting the breakup cross
sections with the respective experimental data (see, e.g.
\cite{Ingemarsson2002}) to make conclusions on the contributions
of the breakup channel to the whole picture of scattering. In
fact, it is seen in Fig.~\ref{fig7} a broad minimum of the
Im$U^{\prime}_{opt}(r)$ around $r$=2.6 fm which illustrates
qualitatively the strong effect of the breakup channel on the
elastic scattering cross section.

\section{SUMMARY AND CONCLUSIONS \label{s:concl}}

The results of the present work can be summarized as follows:

i) The optical potentials and cross sections of $^{8}$He+p elastic
scattering were calculated at the energies of 15.7, 26.25, 32, 66
and 73 MeV/N and comparison with the available experimental data
was performed.

(a) The direct and exchange parts of the real OP ($V^F$) were
calculated microscopically using the folding procedure and density
dependent M3Y (CDM3Y6-type) effective interaction based on the
Paris NN potential.

(b) The imaginary part of the OP ($W^H$) was calculated using the
high-energy approximation.

(c) Three different model densities of protons and neutrons in
$^{8}$He were used in the calculations: the Tanihata densities
\cite{Tani92}, the LSSM densities \cite{Kara2000,Amos2000b} and
the densities obtained in an approach \cite{Massen99,Moust2000}
with accounting for the central-type NN short-range Jastrow
correlations.

(d) The spin-orbit contribution to the OP was also included in the
calculations.

(e) The $^{8}$He+p elastic scattering differential cross sections
and total reaction cross sections were calculated using the
program DWUCK4 \cite{Kunz93}.

ii) The density and energy dependence of the effective NN
interaction were studied. It was shown that the behavior of the
cross sections for given energies and interval of angles is weakly
sensitive to the choice of the model for the $^{8}$He density. The
further calculations of the cross sections were performed using
the LSSM density since it has a realistic exponential behavior in
a peripheral region of the nucleus.

iii) It was shown that the effects of the Jastrow central
short-range NN correlations on the OP's and on the shape of
differential cross sections are weak.

iv) We note that the regularization of the OP's used in this work
by the introduction of the fitting parameters ($N$'s) can serve as
a quantitative test of our method, but not as a tool to obtain a
best agreement with the experimental data. The problem of the
ambiguity of the values of the parameters $N_R$, $N_I$,
$N_{R}^{SO}$, $N_{I}^{SO}$ (that give the ``depth'' of each
component of the OP) when the fitting procedure is applied to a
limited number of experimental data is considered. It was shown
that, generally, at energies $E>20$ MeV/N a good agreement with
the experimental data for the differential cross sections can be
achieved by varying mainly the volume part of the OP neglecting
the SO contribution. A physical criterion imposed in our work on
the choice of the values of the parameters $N$ was the known
behavior (e.g. \cite{Romanovsky}) of the volume integrals $J_{V}$
and $J_{W}$ as functions of the incident energy in the interval
$0<E_{inc}<100$ MeV/N. Another criterion is related to the values
of the total cross section of scattering and reaction. However,
the corresponding empirical data for these values are missing at
the energy interval considered in our work.

v) It was shown that the difficulties arising in the explanation of
the $^{8}$He+p cross sections at lower energies (e.g.~15.7 and 26.25
MeV/N) lead to the necessity to account for the effects of the
nuclear surface (and, correspondingly, of the diffuse region of the
OP). For this reason we included in the cross section calculations
the surface component of the OP and applied it to the case of
$E$=15.7 MeV/N. In our opinion, the account of the latter can be
considered as an imitation of the breakup channel effects. A more
successful explanation of the cross section at low energies could be
given by inclusion of polarization contributions due to virtual
excitations of inelastic and decay channels of the reactions.

\begin{acknowledgments}
The work is partly supported on the basis of the Project from the
Agreement for co-operation between the INRNE-BAS (Sofia) and JINR
(Dubna) and from the Agreement between BAS and Aristotle
University of Thessaloniki. Three of the authors (D.N.K., A.N.A.
and M.K.G.) are grateful for the support of the Bulgarian Science
Fund under Contracts Nos.~02--285 and $\Phi$--1501. The authors
E.V.Z. and K.V.L. thank the Russian Foundation for Basic Research
(Grant No. 09-01-00770) for the partial support.
\end{acknowledgments}

\end{document}